\documentclass[aps,epsfig,twocolumn]{revtex4}

\usepackage{graphicx} 
\usepackage{dcolumn} 

\usepackage{amssymb}
\usepackage{amsmath}
\usepackage{bm}
\usepackage[dvips]{color}

\newcommand{\bq}{\begin{equation}} 
\newcommand{\eq}{\end{equation}}
\newcommand{\bqa}{\begin{eqnarray}} 
\newcommand{\eqa}{\end{eqnarray}}
\newcommand{\nn}{\nonumber \\}
\newcommand{\bw}{\begin{widetext}}
\newcommand{\ew}{\end{widetext}}

\begin{document}
\title{ Stability of the U(1) spin liquid with spinon Fermi surface in 2+1 dimensions }

\author{Sung-Sik Lee}
\affiliation{Department of Physics $\&$ Astronomy, McMaster University,              Hamilton, Ontario L8S 4M1, Canada}

\date{\today}

\begin{abstract}
We study the stability of the 2+1 dimensional U(1) spin liquid state against proliferation of instantons in the presence of a spinon Fermi surface.
By mapping the spinon Fermi surface into an infinite set of 1+1 dimensional chiral fermions,
it is argued that an instanton has an infinite scaling dimension for any nonzero number of spinon flavors.
Therefore, the spin liquid phase is stable against instantons and the non-compact U(1) gauge theory is a good low energy description.
\end{abstract}
\maketitle

\section{Introduction}
Fractionalized phase is a novel state
of correlated many-body systems
where low energy excitations carry fractional quantum numbers
of microscopic degrees of freedom.
In 1+1D, a spin-charge separation, which is an example of fractionalization, can naturally
occur due to the low dimensionality\cite{HALDANE}.
In 2+1D, fractional quantum Hall states support
excitations which have fractional electric charges\cite{LAUGHLIN}.
Finding a fractionalized phase
in time-reversal symmetric 2+1D systems
is an outstanding problem
in condensed matter physics\cite{ANDERSON,PALEE}.

In fractionalized phases, 
there exist non-local correlations 
which are not captured by 
the conventional symmetry breaking picture\cite{Wen_SL}.
Those correlations are associated with 
a condensation of stringy objects in space\cite{WEN} 
or membranes in space-time\cite{LEE}. 
It turned out that 
the most natural framework 
to describe those correlations 
is gauge theory, where the gauge field
describes transverse fluctuations 
of condensed strings or membranes.

Spin liquid is a fractionalized state where 
an elementary excitation is spinon 
which carries spin $1/2$ but no charge\cite{ANDERSON}.
Among a variety of possible spin liquid states\cite{Wen_SL}, 
the state which has fermionic spinons and an emergent U(1) gauge field
has been proposed for many 2+1D strongly correlated electron systems including
high temperature superconductors, frustrated magnets and heavy fermion systems.
Although high $T_c$ superconductors have the conventional 
superconducting ground state, 
the normal state shows non-Fermi liquid behaviors 
which are possibly due to a proximity to a spin liquid state\cite{PALEE,SENTHIL_U1}. 
Frustrated magnets are simpler systems than the high $T_c$ cuprates
in that there is no low energy charge mode.
At the moment, there are promising candidate materials\cite{SHIMIZU1,HELTON} 
for which the U(1) spin liquid states with fermionic spinons have been proposed\cite{MOTRUNICH,Lee_U1,RAN}.
Related fractionalized phases have been studied 
in heavy fermion systems near magnetic quantum critical points\cite{SenthilVojtaSachdev}
and frustrated bose systems\cite{MotrunichFisher}.
In the U(1) spin liquid states,
fermionic spinons have either nodal points or Fermi surfaces.
The gapless spinons are strongly coupled with the U(1) gauge field at low energies and
there is no well-defined quasiparticle\cite{RANTNER,PLEE92,HALPERIN,POLCHINSKY,KIM94,ALTSHULER}.

Because of underlying lattice structures, 
the U(1) gauge field is compact, 
which allows for a topological defect called instanton (or monopole).
Instanton, as a localized object in space-time, 
describes an event where the flux of the gauge field changes by $2 \pi$.
Understanding dynamics of instantons is crucial 
because the fractionalized state is stable 
only if instantons are suppressed in the long distance limit.
It has been known that if there is no gapless spinon,
instantons always proliferate, resulting in confinement.
In this case, spin liquid states are not stable 
and spinons are permanently confined\cite{POLYAKOV77}.
In the presence of gapless spinons, it is possible that
the gauge field is screened and instanton becomes irrelevant
in the low energy limit.
If this happens, fractionalized phase is stable 
and spinons arise as low energy excitations.

If there are a large number of gapless spinons 
which have the relativistic dispersion near nodal points, 
it has been shown that instanton is irrelevant at low energies
and the fractionalized phase is stable\cite{HERMELE}.
However, it is largely unknown whether the spin liquid phase
is stable in the physical cases where the number of spinon flavors is relatively small.
In the presence of spinon Fermi surface,
it has been speculated that the abundance of low energy spinon modes 
may stabilize the fractionalized phase more easily.
However, dynamics of instantons 
in the presence of non-relativistic spinons 
has not been well understood.
There have been several RPA studies\cite{IOFFE,Nagaosa,Ichinose,Herbut,Kim,Kaul}, 
but currently there exists no non-perturbative analysis on the fate of instantons
in the presence of spinon Fermi surface. 
Particularly, a lack of the conformal symmetry makes it hard 
to treat the problem in a non-perturbative way 
which is required because instanton itself is a non-perturbative phenomenon.

In this paper, we provide a non-perturbative argument 
which supports the idea that the U(1) spin liquid state with
spinon Fermi surface is indeed stable 
against proliferation of instantons for any nonzero $N$, 
where $N$ is the number of spinon flavors.
The paper consists of the three parts.
In the first part (Sec. III), we ignore fluctuations of the non-compact component of the 
gauge field and calculate the scaling dimension of instanton 
at the fixed point described by free spinons.
To do this, in Sec. II, we  formulate low energy modes near the Fermi surface
in terms of an infinite number of 1+1D chiral fermions.
Since an instanton is a localized source of $2\pi$ flux in space-time,
the fermions which move in 1+1 dimensional subspaces 
have to enclose the half of the solid angle around the instanton 
and acquire phase $\pi$,
when they are transported around the instanton at a sufficiently large distance.
This is illustrated in Fig. \ref{fig:projection_flux}.
Therefore, an instanton operator corresponds to a twist operator of the 1+1D chiral fermions. 
The scaling dimension of an instanton is infinite because
there are infinitely many 1+1D fermions parametrized by 
the direction of their velocities (or angular momentum), and
each fermion contributes a finite scaling dimension 
to the total scaling dimension of the instanton operator.
In the second part (Sec. IV), the fluctuations of the non-compact gauge field 
are considered together with instantons.
To control the gauge fluctuations, we consider a large $N$ limit.
In this case, vertex corrections are negligible and
we can obtain a definite scaling transformation under which
the low energy theory remains invariant.
The key difference from the previous studies\cite{POLCHINSKY,ALTSHULER}
is that in the present approach all points on the Fermi surface 
are treated on the equal footing rather than focusing on a local patch
in the momentum space.
This enables us to define the scaling dimension of the instanton operator,
taking into account the whole Fermi surface. 
With the fluctuating non-compact gauge field,
fermion modes which have different Fermi velocities
are no longer decoupled, and we can not simply sum 
the scaling dimensions of different modes as we did
in the non-interacting case.
However, in the low energy limit, only small angle scatterings are important 
because momenta of the gauge field are scaled down while
the circumference of the Fermi surface is unchanged under the scale transformation.
This implies that two fermion fields on different points on the Fermi surface
are essentially decoupled at low energies.
Therefore, there are still infinitely many independent 1+1D fermion modes
which contribute to the scaling dimension of instanton at low energies.
By using this property, we can argue that the scaling dimension of an instanton 
is infinite at the interacting fixed point too.
Finally, in Sec. V, we consider the case with a small $N$ of the order of $1$ 
which is directly pertinent to the U(1) spin liquid state 
with two flavors (spin up and down) of spinons\cite{MOTRUNICH,Lee_U1}
proposed for $\kappa-(BEDT-TTF)_2 Cu_2 (CN)_3$\cite{SHIMIZU1}.
With a small $N$, the Fermi surface is strongly coupled with
the fluctuating gauge field and vertex corrections can not be ignored.
This makes it difficult to find an explicit form of a scaling transformation 
for the strongly interacting fixed point.
However, one can see that the essential properties which make the  
scaling dimension of instanton infinite 
does not depend on the specific form of a  scaling transformation.
Actually, the existence of an extended Fermi surface 
and the fact that only small angle scatterings are important at low energies
are enough to argue that 
the scaling dimension of instanton remains to be infinite and 
instantons are irrelevant at the strongly interacting fixed point
for any nonzero $N$.

\section{Angular representation of Fermi surface}

\begin{figure}
        \includegraphics[height=14cm,width=8cm]{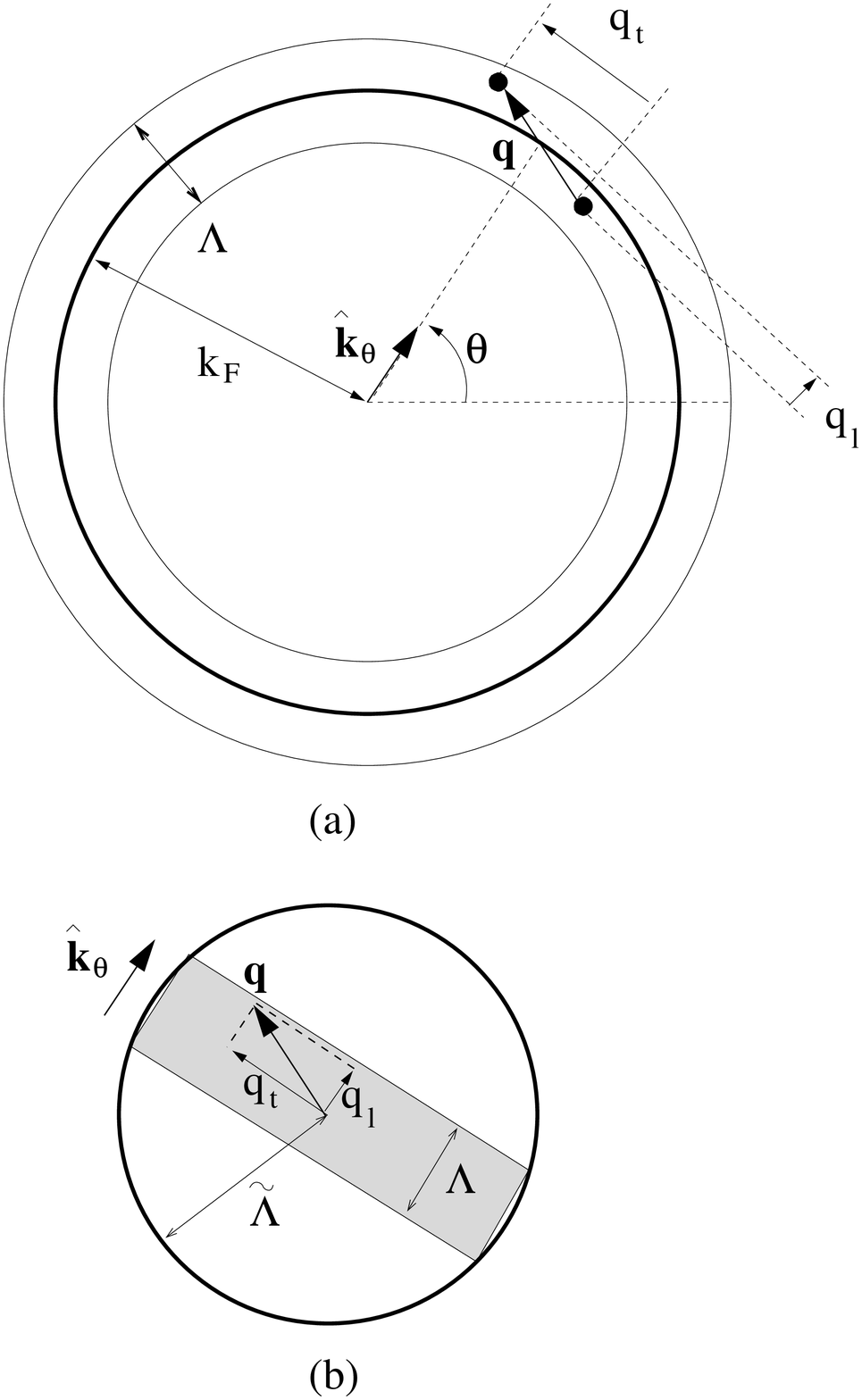} 
\caption{ 
Low energy modes in the momentum space (a) for the fermions 
and (b) for the gauge field.
$\Lambda$ and $\tilde \Lambda$ are the cut-off momenta for the fermions and the gauge field respectively. 
In (a), the bold circle represents the Fermi surface 
and the arrow connecting the two filled circles represents
a momentum transfer from the gauge field.
The shaded strip in (b) represents the points of momenta which 
are included in the background gauge field felt by the 1+1D chiral fermions.
}
\label{fig:FS}
\end{figure}

We start by considering $N$ flavors of fermions coupled with a compact U(1) gauge field in 2+1D,
\bqa
S & = & \int d^3 x \Bigl[
\Psi_j^* (  \partial_0 - i a_0 - \mu_F ) \Psi_j \nn
&& + \frac{1}{2m} \Psi_j^* ( -i {\bf \nabla } -  {\bf a})^2 \Psi_j 
 + \frac{1}{4g^2} f_{\mu \nu} f_{\mu \nu}
\Bigr].
\eqa
Here $\Psi_j$ is the fermion field with $N$ flavors, $j=1,2,..,N$
and $a_\mu = (a_0, {\bf a})$ is the U(1) gauge field with $\mu=0,1,2$.
$\mu_F$ is the chemical potential and
$g$, the gauge coupling.
$f_{\mu \nu}$ is the field strength tensor.
Summation over the repeated flavor index $j$ is implied.
In the energy-momentum space, the action becomes
\bw
\bqa
&& S  =  \int d^3p ~~ 
\Bigl[ ( i p_0 + \epsilon_{\bf p} ) 
\Psi_{j}^*(p) \Psi_{j}(p)  
+ \frac{1}{2g^2} \left(
p^2 \delta_{\mu \nu} - p_\mu p_\nu \right)
a_\mu^*(p) a_\nu(p) \Bigr] \nn
& & +  \int \frac{d^3p d^3l}{(2\pi)^{3/2}} ~~ 
\left( -i a_0(l) - \frac{{\bf p} \cdot {\bf a}(l)}{m} \right) \Psi_j^*(p+\frac{l}{2}) \Psi_{j}(p-\frac{l}{2}) \nn
&& +  \int \frac{d^3p_1 d^3p_2 d^3l}{(2\pi)^{3}} ~~\frac{1}{2m} {\bf a}^*(p_2-l) \cdot {\bf a}(p_2)
\Psi_{j}^*(p_1+l) \Psi_j(p_1).\nn
\eqa
Here $p$, $l$ denote energy-momentum vectors and $\epsilon_{\bf p} = \frac{|{\bf P}|^2}{2m} - \mu_F$.
Integrating out high energy fermion modes outside a momentum shell with a width $\Lambda$ 
near the Fermi surface, 
we obtain the low energy effective action $S=S_0+S_1$, where
\bqa
&& S_0   =  \int d \omega  d k  d \theta ~~
\left( i\omega + k \right)  
\psi_{j}^*(\omega, k, \theta ) \psi_{j}(\omega, k, \theta) \nn
&&
 +  \int d^3p  
\left( 
\frac{1}{2g^2} \left[
 p^2 \delta_{\mu \nu} - p_\mu p_\nu \right]
a_\mu^*(p) a_\nu(p)
+  K  {\bf a}^*(p) \cdot {\bf a}(p)  \right), \nn
&& S_1 =  - \frac{i}{( 2\pi)^{3/2}} 
\int d \omega 
 d k 
 d \theta
 d \nu 
 d q_{l} 
 d q_{t}  ~
\Bigl(
 a_0(\nu,q_l,q_t;\theta) -i a_\theta(\nu,q_l,q_t;\theta)
\Bigr) \times  \nn
&&
 \psi_{j}^*(\omega+\frac{\nu}{2}, k+\frac{q_{l}}{2}, \theta+ \frac{q_t}{2k_F})
 \psi_{j}(\omega-\frac{\nu}{2}, k-\frac{q_{l}}{2}, \theta- \frac{q_t}{2k_F}).\nn
\label{6}
\eqa
\ew
Here, the Fermi velocity has been set to $1$.
Fermion momentum is represented in the polar coordinate\cite{Shankar} where
$k \equiv |\textbf{k}| - k_F$ is the deviation of momentum from the Fermi surface in the radial direction
and $\theta$ is the angular coordinate as is shown in Fig. \ref{fig:FS} (a).
We use the approximation,
$\int d\textbf{k} = \int d |\textbf{k}| |\textbf{k}| \int d \theta \approx k_F \int dk \int d\theta$ 
and redefine the fermion field as 
$\psi_{j}(\omega,k,\theta) \equiv k_F^{1/2} \Psi_{j} \left(\omega,k_1=( k_F + k ) \cos\theta, k_2= (k_F + k) \sin\theta \right)$. 
$K \sim N k_F$ is the diamagnetic term.
$a_\theta = \hat k_\theta \cdot {\bf a}$ is the spatial gauge field 
parallel to the fermion momentum along $\hat k_\theta = ( \cos \theta, \sin \theta)$.
$q_l=\hat k_\theta \cdot {\bf q}$ and $q_t=( \hat k_\theta \times {\bf q})_z$ are 
the momentum components of the gauge field 
which are parallel and perpendicular to $\hat k_\theta$ respectively.
Note that $a_0(\nu,q_l,q_t;\theta)$ and $a_\theta(\nu,q_l,q_t;\theta)$ in the second line of Eq. (\ref{6}) 
implicitly depend on $\theta$
because $q_l$ and $q_t$ are measured with reference to $\hat k_\theta$ as is shown in Fig. \ref{fig:FS} (b).
$\Lambda$ is the momentum cut-off of the fermions near the Fermi surface
and $\tilde \Lambda$ is the  cut-off of the gauge field. 
For $\Lambda << k_F$, we can ignore the quadratic term $k^2/2m$ which is irrelevant at low energies.

\section{Free fermions}

The gauge field can be decomposed into 
the singular part which includes instanton configurations and 
the non-singular part which describes fluctuations of the non-compact component.
First, we ignore the non-compact gauge field 
and examine the effect of instantons on the free fermions.
For this, we consider a background gauge field $a_\mu^s$ generated by an instanton 
located at $\tau=0$ and $\textbf{x}=0$ in space and time.
For the singular part of the gauge field, we use the temporal gauge where $a_0^s=0$.

\begin{figure}[h!]
        \includegraphics[height=14cm,width=8cm]{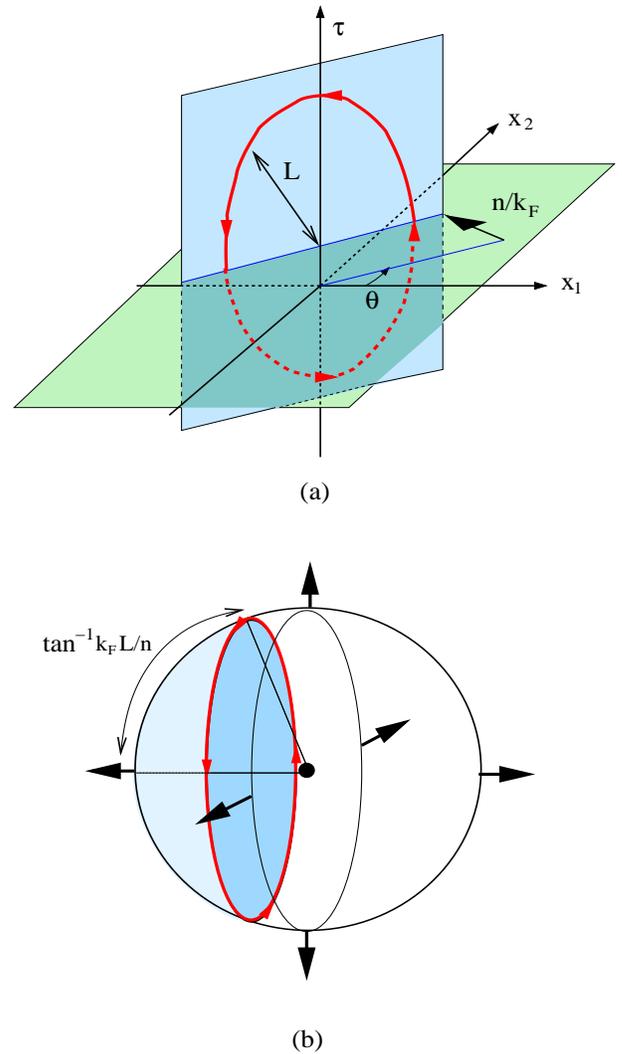} 
\caption{ 
A 1+1D chiral fermion which has an angular momentum $n$ is coupled to a background gauge field
projected to the plane which is represented as the vertical plane in (a).
The arrowed circle in (a) is a trajectory of a fermion transported around an instanton at distance $L$
and the arrowed circle in (b) is the trajectory of the fermion projected on the unit sphere.
The area of the shaded region in (b) is the solid angle enclosed by the fermion around the instanton which is represented as the black dot at the center of the sphere.
In the large distance limit ($L>> n/l_F$), the fermion acquires phase $\pi$ as it moves around the instanton.
}
\label{fig:projection_flux}
\end{figure}

If the Fermi surface has the rotational symmetry,
the field strength for single instanton centered at the origin
has the rotational symmetry too.
This enables us to choose 
$a_\theta^s( \nu,q_l,q_t;\theta)$
to be independent of $\theta$.
Because of the rotational symmetry, 
it is convenient to introduce 1+1D chiral Fermi fields
which have good angular momentum quantum number,
\bq
\psi_{jn}(\tau,x) = \frac{1}{(2 \pi)^{3/2}} 
\int d\omega  d k  d\theta ~
e^{i (\omega \tau + k x + n \theta)} \psi_{j}(\omega, k, \theta).
\label{ar}
\eq
In the 1+1D real space, the action for the chiral fermions becomes
\bqa
S  & = &  \sum_n
\int d \tau d x ~
 \psi_{jn}^*(\tau, x)
\Bigl[
 \partial_\tau 
\nn
&& 
~~~ - i ( \partial_x - i a_\theta^s(\tau,x,x_t = n/k_F) ) 
\Bigr]
\psi_{jn}(\tau, x), 
\label{9}
\eqa
where the 1+1D gauge field `felt' by the chiral fermions is given by 
\bqa
a^s_\theta(\tau,x,x_t) & = & 
\frac{1}{(2 \pi)^{3/2}} 
\int_{-\infty}^\infty d\omega 
\int_{-\Lambda/2}^{\Lambda/2} d q_l 
\int_{-\tilde \Lambda}^{\tilde \Lambda} d q_t \nn
&& e^{i (\omega \tau + q_l x + q_t x_t )} a_\theta^s(\omega, q_l, q_t;\theta).
\label{gm}
\eqa
Here $x$ and $x_t$ represent the displacements 
from the instanton 
in the directions parallel and perpendicular to $\hat k_\theta$ for some $\theta$ respectively.
Because of the rotational symmetry, $a_\theta(\tau,x,x_t)$ is independent of $\theta$
and the choice of $\theta$ does not matter.
The action in Eq. (\ref{9}) describes an infinite set of 1+1D chiral fermions 
coupled to the background gauge field,
 $a_\theta^s(\tau,x,x_t = n/k_F)$.
The chiral fermion with angular momentum $n$ `sees' the 2+1D gauge field 
projected onto a plane 
which is perpendicular to the $x_1-x_2$ plane 
and shifted by $n/k_F$ away from the origin 
as is shown in Fig. \ref{fig:projection_flux} (a).
The range of the momentum integration in Eq. (\ref{gm}) 
is restricted to be within the strip with the width $\Lambda$ 
as is shown in Fig. \ref{fig:FS} (b) and
$a^s_\theta(\tau,x,x_t)$ represents slowly varying configurations of the gauge field 
in space and time.
Nevertheless, the components with large momenta 
become unimportant in the long distance limit,
and $a^s_\theta(\tau,x,x_t)$ accurately describes 
the true configuration of instanton 
far away from the center.
For an instanton whose field strength is isotropic in space and time, 
the gauge field is given by
\bqa
a_0 & = & 0, \nn
a_1 & = & \frac{ x_2}{2 r ( r- \tau)},\nn
a_2 & = & -\frac{ x_1}{2 r ( r- \tau)},
\label{igf}
\eqa
with $r=\sqrt{ \tau^2 + x_1^2 + x_2^2 }$.
In this gauge, there is a Dirac string stretched along the
positive $\tau$ axis.
The presence of the Dirac string is not important because
the infinitely thin tube of $2 \pi$ flux can always be placed
inside a halo of an underlying lattice and the unit flux can be
gauged away.
The gauge field in Eq. (\ref{igf}) represents 
an instanton with the Lorentz symmetry.
In the presence of the non-relativistic fermions, 
the space-time isotropy is lost and the field strength 
will be redistributed.
With the broken Lorentz symmetry, the spatial rotational symmetry in the $x_1-x_2$ 
space may or may not be broken.
In the following, we will first consider the case with the spatial rotational symmetry
and then consider general cases without the symmetry.

As an 1+1D fermion is transported within a plane at distance $L$ from its origin as in Fig. \ref{fig:projection_flux} (a), 
it encloses the solid angle, 
$\Omega_n(L) = 2\pi \left( 1 - \frac{n}{ \sqrt{(L k_F)^2 + n^2}} \right)$ 
in the unit sphere around the instanton as is shown in Fig. \ref{fig:projection_flux} (b).
Since an instanton is the source of flux $2 \pi$,
the fermion acquires a non-trivial phase $\Phi_n(L)$.
Without loss of generality, we can defined the phase angle within the interval $(-\pi,\pi]$.
For the Lorentz symmetric instanton configuration in Eq. (\ref{igf}), 
the phase angle is the half of the solid angle and becomes
\bq
\Phi_n(L) = sgn(n) \pi \left( 1 - \frac{|n|}{ \sqrt{(L k_F)^2 + n^2}}  \right).
\label{angle}
\eq

For non-isotropic configurations, $\Phi_n(L)$ will be different from Eq. (\ref{angle}).
However, the explicit form of $\Phi_n(L)$ is not important for the following discussions.
What is important is the fact that $|\Phi_n(L)| \rightarrow \pi$ for any finite $n$ as $L \rightarrow \infty$
in the presence of the spatial rotational symmetry.
This is because trajectories of fermions projected onto the unit sphere around the instanton
will eventually follow a big circle for any angular momentum $n$ in the large $L$ limit, 
and the flux enclosed by any half sphere that cuts through the north and south poles is always $\pi$ 
due to the spatial rotational symmetry.
Therefore, in the long distance limit, an instanton twists boundary conditions of all fermions
from the periodic condition to the anti-periodic one.
Roughly speaking, at a length scale $L$, 
an instanton twists boundary conditions of the fermions 
which have angular momenta $|n| < k_F L$ by $\pi$.

Having understood that an instanton operator corresponds to a twist operator,
we can determine the scaling dimension of instanton.
To do this, we represent the 1+1D space in terms of a complex variable,
$z = \tau - i x$ and rescale the fermion fields to write the action in the standard form,
\bqa
S_{free} = \frac{1}{2 \pi} \sum_{j=1}^N \sum_{n=-\infty}^\infty \int dz d {\bar z}~~ \psi_{jn}^* {\bar \partial} \psi_{jn}, 
\eqa
where $\bar \partial = \frac{\partial}{\partial \bar z}$.
This free theory is invariant under the scale transformation,
\bqa
z & = & b z^{'}, \nn
\bar z & = & b \bar z^{'}, \nn
\psi_{jn}(b \tau^{'}, b x^{'}) & = & b^{-1/2} \psi_{jn}^{'} (\tau^{'}, x^{'})
\label{scale1}
\eqa
with $b>1$.
With instantons, the free theory is perturbed as
\bqa
S = S_{free} +  y \int dz d \bar z~~ \sigma,
\eqa
where $\sigma$ is the creation operator of an instanton or an anti-instanton
and $y$ is the fugacity.
Since an instanton twists the boundary conditions 
of all fermions in the low energy limit,
the instanton operator can be written as
\bqa
\displaystyle \sigma = \Pi_{j,n} \sigma_{jn}(\pi),
\eqa
where $\sigma_{jn}(\pi)$ is the operator which twists the boundary condition of $\psi_{jn}$ by $\pi$.
However, the scaling dimension of $\sigma$ is not well-defined because
the operator ends up twisting the infinite number of fermions by the finite angle, $\pi$.
Actually, not all fermions are twisted by the same angle at a finite length scale $L$\cite{NONLOCAL}.
Therefore we introduce a regularized instanton operator, 
\bqa
\displaystyle \sigma_L = \Pi_{j,n} \sigma_{jn}(\Phi_n(L)),
\eqa
where $\sigma_{jn}(\Phi_n(L))$ is an operator which twists the boundary condition of 
$\psi_{jn}$ by the angle $\Phi_n(L)$. 
Physically, $\sigma_L$ creates the flux configuration near an instanton at a length scale $L$.
Although $\sigma_L$ is not a true instanton operator, 
we can learn about the property of the true instanton 
by taking $L \rightarrow \infty$ limit of $\sigma_L$.
The point of introducing the regularized operator is that $\sigma_L$ is a well defined local operator 
which has a finite scaling dimension, as will be shown below.

Since the scaling dimension corresponds to the eigenvalue of the scale transformation
generated by the Noether current $j_\nu = x^\mu T_{\mu \nu}$, 
the scaling dimension $d_L$ of the regularized instanton operator can be obtained from
\bqa
d_L \sigma_L(0) = \oint  \frac{dz}{2\pi i}~~ z T(z) \sigma_L(0),
\label{c}
\eqa
where $T(z)$ is the holomorphic energy momentum tensor.
In the state-operator correspondence, we can view the scaling dimension 
as the `energy' of the quantum state defined on the circle around the origin
associated with the `time' evolution in the radial direction.
What the insertion of the $\sigma_L$ operator does is 
to twist the boundary condition of $\psi_{jn}$ by $\Phi_n(L)$.
Then we can rewrite Eq. (\ref{c}) as
\bqa
d_L =   \left< \oint   \frac{dz}{2\pi i}~~ z T(z) \right>_{twisted ~ b.c.},
\label{c2}
\eqa
where we impose the twisted boundary condition for the fermion fields around the origin.
Each fermion $\psi_{jn}$ contribute a scaling dimension 
$d(\Phi_n(L)) = \frac{\Phi_n(L)^2}{8 \pi^2}$ (see appendix A for derivation) 
and the total scaling dimension for the regularized instanton operator becomes
\bqa
d_L & = &   \sum_{j=1}^N \sum_{n=-\infty}^\infty \frac{\Phi_n(L)^2}{8 \pi^2}.
\label{D}
\eqa
Since $\Phi_n(L)$ approaches $\pi$ as $L$ increases,
$d_L$ diverges in the large $L$ limit.
For the isotropic instanton configuration, we have
\bqa
d_L^{isotropic} & = &   \sum_{j=1}^N \sum_{n=-\infty}^\infty \frac{1}{8}  \left( 1 -  \frac{|n|}{ \sqrt{(L k_F)^2 + n^2}}  \right)^2,
\label{D2}
\eqa
and it is easy to check that it diverges linearly with $L$.
The renormalization group equation for the fugacity becomes 
\bqa
\frac{d y_L}{d \ln b} = \left( 2 - d_L \right) y_L + O(y_L^2),
\label{flow}
\eqa
where $y_L$ is the fugacity of the regularized instanton operator.
The present approach does not allow us to calculate the higher order terms in $y$. 
However, from the linear term alone, we can readily see that
for any $N>0$ the regularized instanton operator for a sufficiently large $L$ 
(hence the true instanton operator defined as $\sigma = \lim_{L \rightarrow \infty} \sigma_L$) 
is strongly irrelevant at the fixed point with $y=0$.
Namely, a small nonzero $y$ will flow to the fixed point with $y=0$.
Although the scaling dimension of the true instanton operator defined as $d = \lim_{L\rightarrow \infty} d_L$
is ill-defined (infinite) for any $N>0$, 
the regularized scaling dimension diverges more rapidly with increasing $L$ when $N$ is larger.
This is consistent with the physical intuition that the presence of more fermions
results in a larger scaling dimension of instanton via screening.

Until now, we have considered the case with the spatial rotational symmetry.
If the symmetry is broken by an underlying lattice, 
the field configuration of an instanton is no longer symmetric under the spatial rotation.
Here we will see that the conclusion reached for the rotationally symmetric Fermi surface holds in general cases too
as far as the general Fermi surface can be obtained from the symmetric one through a smooth deformation.
Performing the Fourier transformations for the frequency and the radial momentum, 
we rewrite $S_1$ in Eq. (\ref{6}) as
\bqa
&& S_1 =   - \frac{1}{( 2\pi)^{1/2}} 
\int d \tau 
 d x 
 d \theta
 d \theta^{'}  \nn
&&
  a_\theta(\tau,x,k_F(\theta-\theta^{'});(\theta+\theta^{'})/2)
 \psi_{j}^*(\tau, x, \theta)
 \psi_{j}(\tau, x, \theta^{'}). \nn
\eqa
For a general Fermi surface, $a_\theta$ has a non-trivial angular dependence on $(\theta+\theta^{'})/2$. 
To diagonalize the action, we have to use a different basis than the angular momentum basis.
We consider the basis transformation,
\bq
\psi_{jn}(\tau,x) = \frac{1}{(2 \pi)^{1/2}} \int d\theta ~
f_n^*(\tau,x,\theta) \psi_{j}(\tau, x, \theta),
\label{ar2}
\eq
where $f_n(\tau,x,\theta)$ satisfies the eigenvalue equation,
\bqa
&& \frac{1}{(2\pi)^{1/2}} \int d \theta^{'} a_\theta(\tau,x,k_F(\theta-\theta^{'});(\theta+\theta^{'})/2) f_n(\tau,x,\theta^{'}) \nn
&& = a_n(\tau,x) f_n(\tau,x,\theta)
\eqa
and the normalization condition
\bqa
\frac{1}{2 \pi} \int d\theta ~
f_m^*(\tau,x,\theta) f_n(\tau,x,\theta) = \delta_{m,n}.
\eqa
One can always find such a basis because the kernel $a_\theta$ satisfies the Hermitian condition,
$a_\theta(\tau,x,k_F(\theta-\theta^{'});(\theta+\theta^{'})/2) = a_\theta^*(\tau,x,-k_F(\theta-\theta^{'});(\theta+\theta^{'})/2)$ at each $\tau$ and $x$.
In the new basis, the action for the 1+1D chiral fermions becomes diagonal,
\bqa
S  & = &  \sum_n
\int d \tau d x ~
\psi_{jn}^*(\tau, x)
\Bigl[
 \partial_\tau \nn
&& 
- i ( \partial_x - i a_n(\tau,x) ) 
\Bigr]
\psi_{jn}(\tau, x). 
\label{92}
\eqa
The phase acquired by the $n$-th fermion when the fermion is transported around the instanton is 
\bqa
\Phi_n = \oint dx a_n,
\eqa
and the total scaling dimension becomes 
\bqa
d = N \sum_{n} \frac{\Phi_n^2}{8 \pi^2},
\eqa
where $-\pi< \Phi_n \leq \pi$.
For the rotationally symmetric Fermi surface, the index $n$ represents the angular momentum as before
and $\Phi_n = \pi$ for all $n$ in the low energy limit.
As the Fermi surface (hence the field configuration of an instanton) is distorted,
the distribution of $\Phi_n$ gets broadened around $\pm \pi$ and $d$ will decrease.
However, $d$ can not change abruptly as the Fermi surface is smoothly deformed
and $d$ will remain infinite in the thermodynamic limit unless there is a phase transition
associated with the topology of the Fermi surface.
Therefore, the scaling dimension of an instanton is infinite for a general Fermi surface
as far as the Fermi surface is smoothly connected to the rotationally symmetric Fermi surface.

\section{Interacting fermions with a large number of flavors}

\begin{figure}[h]
        \includegraphics[height=13cm,width=9cm]{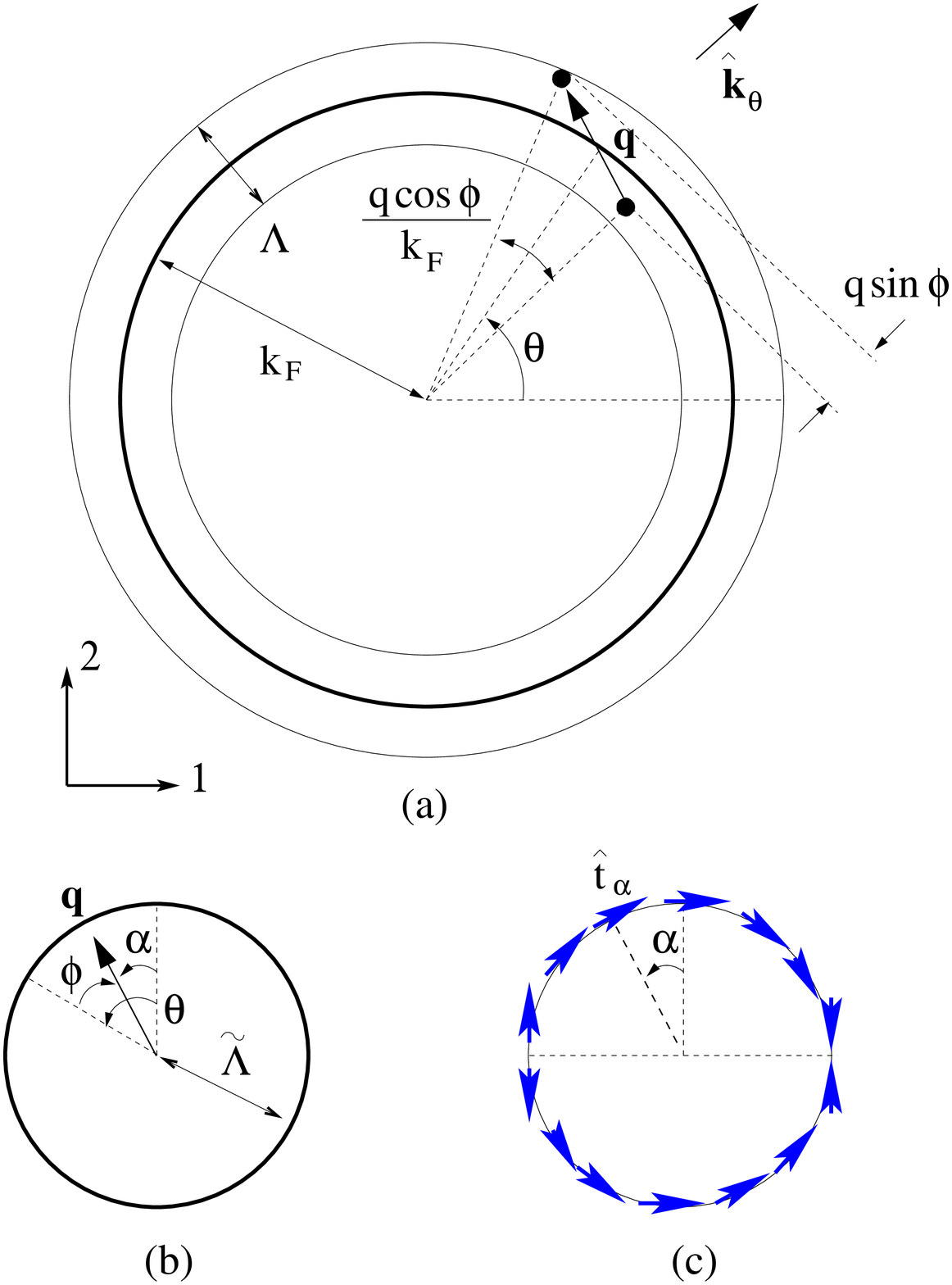} 
\caption{
Angular representations of (a) the fermions and (b) the gauge field.
Note that the angle $\theta$ for the fermions is measured from the $x_1$ axis
and the angle $\alpha$ for the gauge field is measured from the $x_2$ axis
so that the momentum of the gauge field becomes perpendicular to the Fermi momentum
when $\alpha = \theta$.
In (a), the arrow represents a process where a fermion changes its radial momentum and angle 
through an absorption of a momentum from the gauge field.
In (b), $\phi$ parametrizes a deviation of $\alpha$ from $\theta$.
In (c), the arrows represent the polarization vectors of the transverse gauge field
at different points in the momentum space.
 }
\label{fig:FS_angle}
\end{figure}

Now we consider the whole theory by taking into account fluctuations of the non-compact gauge field.
For the non-compact gauge field, 
we will use the Coulomb gauge where ${\bf \nabla} \cdot {\bf a} = 0$\cite{GAUGE}
and drop the temporal gauge field which is screened out at long distances.
In the following, we will focus on the rotationally symmetric case.
With the fluctuating non-compact gauge field, 
the interaction term can be written as
\bw \bqa
&& S_1  =   \frac{-1}{( 2\pi)^{3/2}} 
\int_{-\infty}^\infty d \omega 
\int_{-\frac{\Lambda}{2}}^{\frac{\Lambda}{2}} d k 
\int_{-\pi}^{ \pi} d \theta
\int_{-\infty}^\infty d \nu 
\int_{0}^{\tilde \Lambda} d q q 
\int_{-\pi}^{ \pi} d \phi   
 \nn   && 
~~~ \Theta\left( \Lambda/2 - \left|k +  q \sin \phi/2 \right| \right)
\Theta\left( \Lambda/2 - \left|k -  q \sin \phi/2 \right| \right) 
 \hat k_\theta \cdot \hat t_{\theta-\phi}  
a(\nu,q,\theta-\phi) \times \nn
&&
\psi_{j}^*\left(\omega+ \nu/2, k+ q \sin \phi /2, \theta+ q \cos  \phi/ 2 k_F\right)
\psi_{j}        \left(\omega- \nu/2, k- q \sin \phi /2, \theta- q \cos  \phi/ 2 k_F\right), \nn
\label{52}
\eqa 
where $a$ is the transverse gauge field. 
The momentum of the gauge field is also written  
in the polar coordinate as 
$a(\nu,q,\alpha) = a(\nu,q_1=-q\sin \alpha, q_2=q\cos \alpha )$.
$\hat t_\alpha$ is the polarization vector of the gauge field with an angle $\alpha$,
where the angle is defined in such a way that $\hat t_\alpha$  
becomes parallel to $\hat k_\theta$ when $\alpha = \theta$ 
as is depicted in Fig. \ref{fig:FS_angle}.
We choose the polarization vector as
$\hat t_{\alpha} = ( \cos \alpha, \sin \alpha )$ for $-\pi/2 \leq \alpha < \pi/2$
and $\hat t_{\alpha} = ( \cos (\alpha+\pi), \sin (\alpha+\pi) )$ for $\alpha \geq \pi/2$ or $\alpha < -\pi/2$
as is shown in \ref{fig:FS_angle} (c).
In this definition, there are discontinuities in $\hat t_{\alpha}$ at $\alpha = \pm \pi/2$.
However, this definition is convenient to make the reality of the gauge field explicit in 
the momentum space as $a(-\nu,q,\alpha+\pi) = a^*(\nu, q,\alpha)$.
$\Theta(x)$ is the step function which ensures that
the momenta of fermions lie 
within the shell of the width $\Lambda$ near the 
Fermi surface.

At low energies we have $\tilde \Lambda \sim \sqrt{ k_F \Lambda}$
because the Fermi surface is locally parabolic.
Since $k_F >> \tilde \Lambda >> \Lambda$, typical momenta of the gauge 
field are perpendicular to fermion momentum 
and much larger than $\Lambda$. 
Therefore the support of the $\phi$ integration in Eq. (\ref{52})
is sharply centered at $\phi = 0$ and $\phi = \pi$ 
with the width of the order of $\Lambda/\tilde \Lambda << 1$. 
This allows us to use $\sin \phi \approx \pm \phi$ 
and $\cos \phi \approx \pm 1$ in the low energy limit.
Because both of the fermions with angles $\theta$ and $\theta+\pi$ are coupled with 
both of the gauge fields with angle $\theta$ and $\theta + \pi$, 
it is convenient to define two separate fields 
for the opposite-moving fermions
and restrict the integrations of $\theta$ to run from $-\pi/2$ to $\pi/2$.
At the same time, we allow $q$ to run from $-\tilde \Lambda$ to $\tilde \Lambda$ 
and restrict the $\phi$ integration to $-\Lambda/|q| < \phi < \Lambda/|q|$
to implement the theta functions in Eq. (\ref{52}).
Then  the action can be written as $S=S_0 + S_1$, where
\bqa
&& S_0  =  \sum_{s=\pm 1} 
\int_{-\infty}^\infty d \omega 
\int_{-\Lambda/2}^{\Lambda/2} d k 
\int_{-\pi/2}^{\pi/2} d \theta~~
\left( i \omega + s k \right) 
\psi_{js}^*\left(\omega, k,\theta \right)
\psi_{js}\left(\omega, k,\theta \right) 
\nn
&&  + \int_{-\infty}^\infty d \nu 
\int_{-\tilde \Lambda}^{\tilde \Lambda} d q |q| 
\int_{-\pi/2}^{\pi/2} d \alpha~~
\left( 
\frac{1}{2g^2} 
( \nu^2 + q^2 ) 
+  K  \right) a(\nu,q,\alpha) a(-\nu,-q,\alpha), 
\label{53}
\\ 
&& S_1 = - \sum_{s=\pm 1} \frac{1}{( 2\pi)^{3/2}} 
\int_{-\infty}^\infty d \omega 
\int_{-\Lambda/2}^{\Lambda/2} d k 
\int_{-\pi/2}^{ \pi/2} d \theta
\int_{-\infty}^\infty d \nu 
\int_{-\tilde \Lambda}^{\tilde \Lambda} d q |q| 
\int_{- \Lambda / |q|}^{ \Lambda/|q|} d \phi  \nn
&& s a(\nu,q,\theta-\phi)
\psi_{js}^*\left(\omega+\nu/2,  k+  q \phi/2,\theta+s  q/2k_F\right)
\psi_{js}        \left(\omega-\nu/2,  k-  q \phi/2,\theta-s  q/2k_F\right), \nn
\label{54}
\eqa
\ew
where $s=1, -1$ labels fermion fields on the two sides of the Fermi surface at each angle $-\pi/2 \leq \theta < \pi/2$, defined as
\bqa
\psi_{j+}(\omega, k, \theta) & = &  \psi_{j}(\omega, k, \theta) \nn
\psi_{j-}(\omega, k, \theta) & = &  \psi_{j}(\omega, -k, -sgn(\theta) \pi+\theta),
\eqa
and a negative $q$ of the gauge field represents the opposite momentum as
\bqa
a(\nu,-|q|,\alpha) = a(\nu,|q|,\alpha+\pi)
\eqa
with $-\pi/2 \leq \alpha < \pi/2$.
$\psi_{j+}$ and $\psi_{j-}$ have the opposite velocities and they form a two-component 1+1D Dirac fermion.

\begin{figure}[h]
        \includegraphics[height=6cm,width=6cm]{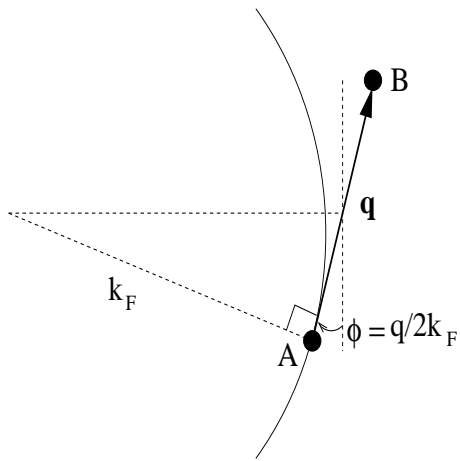} 
\caption{
The effect of the local curvature in the Fermi surface.
As a fermion with a momentum $A$ on the Fermi surface absorbs a momentum ${\bf q}$ 
which is tangential to the Fermi surface at the point,
the fermion ends up with the momentum $B$ which has a higher energy by $q^2/2 k_F$ due to the curvature of the Fermi surface.
This can be seen from Eq. (\ref{54}) as follows.
In order for ${\bf q}$ to be tangential to the Fermi surface at the initial momentum, 
$\phi$ in Eq. (\ref{54}) has to be $q/2k_F$. 
Then the energy of the final state is larger than that of the initial state by 
$q \phi = q^2/2k_F$.
}
\label{fig:FScurvature}
\end{figure}

The present approach is conceptually analogous 
to the bosonized descriptions of Fermi surface\cite{Haldane,Houghton,Castroneto,Kwon}, 
where chiral bosons describe low energy particle-hole excitations near Fermi surface.
However, there is an important difference.
In the previous bosonized description of the Fermi surface coupled with the U(1) gauge field\cite{Kwon}, 
the Fermi surface is taken to be locally flat within each momentum patch, and 
the local curvature is not taken into account.
The present formalism captures the local curvature effect of the Fermi surface, 
which is important to reproduce correct low energy behaviors\cite{ALTSHULER}.
The key is to consider all points on the Fermi surface on the equal footing, 
not treating the Fermi surface as a sum of locally flat Fermi segments.
The way the curvature effect is implemented in Eq. (\ref{54}) is explained in Fig. \ref{fig:FScurvature}.

Here we assume that $N >>1$ in which case the fluctuations of the gauge field are controlled. 
In the leading order of the $1/N$ expansion,
vertex corrections can be ignored\cite{POLCHINSKY}.
The fluctuating gauge field and the gapless fermions lead to 
singular self energies 
and the single particle quantum effective action of the fermions and the transverse gauge field becomes 
(see appendix B)
\bqa
\Gamma_0 & = &  \sum_{s=\pm 1} \int d \omega  d k d \theta 
\left[ i c ~\mbox{sgn}(\omega) |\omega|^{2/3} + s k \right] \nn
&& ~~~~~~~ \psi_{js}^*(\omega, k, \theta ) \psi_{js}(\omega, k, \theta)  \nn
&& + \int  d \nu dq |q|  d \alpha
\left[ \gamma \frac{|\nu|}{|q|} + \chi q^2 \right] a^*(\nu,q,\alpha) a(\nu,q,\alpha), \nn
\label{ga}
\eqa 
where $c$, $\gamma$ and $\chi$ are constants.
Because of the singular quantum corrections, 
the scaling transformation in Eq. (\ref{scale1}) is no longer a symmetry.
Instead, we have to rescale  energy and momentum  as
\bqa
\omega & = & b^{-1} \omega^{'}, \nn
\nu & = & b^{-1} \nu^{'}, \nn
\Lambda & = & b^{-2/3} \Lambda^{'}, \nn
\tilde \Lambda & = & b^{-1/3} \tilde \Lambda^{'}, \nn
k & = & b^{-2/3} k^{'}, \nn
q & = & b^{-1/3} q^{'}.
\label{scale2}
\eqa
Note that the momentum of the fermion in the radial direction 
and the momentum of the gauge field should scale differently.
If we apply this new scale transformation  to $S_1$ in Eq. (\ref{54}), we readily notice that the action can not be made invariant unless the angular variables $\theta$ and $\phi$ are rescaled as well.
This is because momenta of the fermions and the gauge field 
mix with the angular variables through
$k + q \phi$  and $\theta + sq/k_F$.
To make the action invariant, we should assign the scaling dimension $1/3$ to the angular variables.
This is an anomalous scaling dimension of the angular variables which arises solely from quantum effects.
The whole action is invariant if we rescale
\bqa
\theta & = & b^{-1/3} \theta^{'}, \nn
\phi & = & b^{-1/3} \phi^{'}, \nn
\psi_{a}(b^{-1} \omega^{'}, b^{-2/3} k^{'}, b^{-1/3} \theta^{'}) & = & b^{4/3} 
\psi_{a}^{'}(\omega^{'},  k^{'}, \theta^{'}), \nn
a(b^{-1} \nu^{'}, b^{-2/3} q^{'}, b^{-1/3} \phi^{'}) & = & b^{4/3} 
a^{'}(\nu^{'},  q^{'}, \phi^{'})
\label{scale3}
\eqa
along with Eq. (\ref{scale2}).
As we go to lower energy ($b > 1$), 
the range of the $\theta$ integration increases 
from $(-\pi/2,\pi/2)$ to $(-b \pi/2, b\pi/2)$.
In the low energy limit where $b \rightarrow \infty$, $\theta$ becomes a non-compact variable 
which runs from $-\infty$ to $\infty$.
The physical reason behind this `decompactification' of the angular variable can be understood in the following way.
In the low energy limit, the gauge field becomes more and more ineffective in scattering fermions from one momentum to another momentum along tangential directions to the Fermi surface.
This is  because the momentum of the gauge field is scaled down under the scale transformation,
while the circumference of the Fermi surface is unchanged.
This effectively makes two momentum points on the Fermi surface 
more decoupled from each other at lower energies.
In other words, the `metric' of the Fermi surface along the tangential directions 
diverges in the low energy limit compared to the `metric' along the perpendicular directions.

\begin{figure}
        \includegraphics[height=6cm,width=6cm]{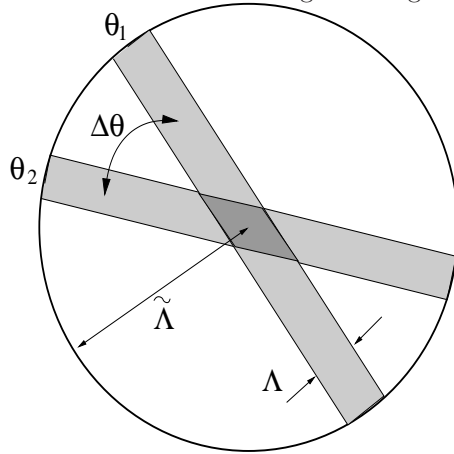} 
\caption{
The shaded areas with angle $\theta_1$ and $\theta_2$ represent momentum points
which are included in $a_q (\tau,x,\theta_1)$ and $a_q (\tau,x,\theta_2)$.
The dark region at the center indicate momentum points which contribute
to both  $a_q (\tau,x,\theta_1)$ and $a_q (\tau,x,\theta_2)$.
The ratio of the area of the dark region to the area of a strip becomes zero in the low energy limit
where the momentum cut-offs scale as $\Lambda \rightarrow b^{-2/3} \Lambda$
and $\tilde \Lambda \rightarrow b^{-1/3} \tilde \Lambda$.
 }
\label{fig:cross}
\end{figure}

Introducing 1+1D fields in real space,
\bw
\bqa
\psi_{js}(\tau,x,\theta) & = & \frac{1}{2\pi} \int_{-\infty}^\infty d\omega \int_{-\Lambda/2}^{\Lambda/2} dk~~ e^{i ( \omega \tau + k x)} \psi_{js}(\omega,k,\theta), \\
\label{10}
a_q (\tau,x,\theta) & = & \frac{1}{2\pi} \int_{-\infty}^\infty d\omega \int_{-\Lambda/|q|}^{\Lambda/|q|} d\phi |q|~~ e^{i ( \omega \tau + x q \phi )} a(\omega,q,\theta- \phi), 
\label{11} 
\eqa
we can write down the low energy effective action in the 1+1D real space as
 \bqa
&& S  = 
 \sum_{s=\pm 1}  
\int d \tau 
d x 
 d \theta~~
\psi_{js}^*\left(\tau, x, \theta \right)
[ \partial_\tau - i s \partial_x ]
\psi_{js}        \left(\tau, x, \theta \right). \nn
&&  + 
\frac{1}{2 \Lambda}
\int d \tau 
 d x 
 d \theta
 d q ~~
|q|
a_q^*(\tau,x,\theta) \left( 
\frac{1}{2g^2} 
( -\partial_\tau^2 + q^2 ) 
+  K  \right) a_q(\tau,x,\theta) \nn
&& - \frac{1}{( 2\pi)^{1/2}} 
\sum_{s=\pm 1}  
\int d \tau 
 d x 
 d \theta
 d q  ~~  
s a_q(\tau,x,\theta)
\psi_{js}^*\left(\tau, x, \theta+s  q/2k_F \right)
\psi_{js}        \left(\tau, x, \theta-s  q/2k_F \right). 
\label{56}
\eqa
\ew
It should be emphasized that in this action $\theta$ is a non-compact variable which runs from $-\infty$ to $\infty$.
In Eq. (\ref{11}), $a_q (\tau,x,\theta)$ depends on four variables while $a(\omega,q,\theta-\phi)$ depends only on three independent variables.
The variable $x$ has been created by the Fourier transformation of $a(\omega,q,\alpha)$ with respect to $\alpha$ which is centered at $\theta$.
Conceptually, this is similar to creating a wave-packet which is localized in both real space and momentum space by linearly superposing wavefunctions whose momenta are centered at a particular momentum.
The factor $1/2\Lambda$ in the second line of Eq. (\ref{56}) is to cancel the double counting of momentum points. 
Note that $a_q (\tau,x,\theta_1)$ and $a_q (\tau,x,\theta_2)$ are not completely independent  for different $\theta_1$ and $\theta_2$.
Both $a_q (\tau,x,\theta_1)$ and $a_q (\tau,x,\theta_2)$ include contributions from a common region in the momentum space as is shown in Fig. \ref{fig:cross}.
However, the overlap is not important in determing the scaling dimension of instanton as will be shown in the following.
The area of the common region for $a_q (\tau,x,\theta_1)$ and $a_q (\tau,x,\theta_2)$ (the dark parallelogram in Fig. \ref{fig:cross}) is 
$( \Lambda^2 / \Delta \theta )$ for a small $\Delta \theta = \theta_2 -\theta_1$.
The ratio of this area to the area included in $a_q (\tau,x,\theta)$ (the long strip in Fig. \ref{fig:cross}) is 
$\gamma = ( \Lambda^2 / \Delta \theta )  / ( \Lambda \tilde \Lambda)$.
As the momentum cut-off decreases  as $\Lambda \rightarrow \Lambda b^{-2/3}$, $\tilde \Lambda \rightarrow \tilde \Lambda b^{-1/3}$, the ratio decreases as $\gamma \rightarrow  b^{-1/3} ( \Lambda / \tilde \Lambda \Delta \theta )$.
The ratio becomes zero in the low energy limit for any nonzero $\Delta \theta$, 
which implies that the two fields which have a finite angle difference (before scaling)
are independent in the low energy limit.
In the rescaled angular variable $\theta^{'} = b^{1/3} \theta$, 
a fixed $\Delta \theta^{'}$ corresponds to a successively reduced $\Delta \theta = b^{-1/3} \Delta \theta^{'}$
as a low energy limit is taken.
Since $\gamma$ goes as $b^{0}$ for a fixed $\Delta \theta^{'}$ in the rescaled variable, 
two fields which have a fixed angle different $\Delta \theta^{'}$ have a finite ratio $\gamma$ in the low energy limit.
On the other hand, two fields whose angle difference increases faster than $\Delta \theta^{'} = b^0$ in the rescaled variable 
have vanishing overlap in the low energy limit.
Since the interval of $\theta^{'}$ increases as $b^{1/3} \pi$, 
there are infinitely many independent fields $a_q (\tau,x,\theta)$ which are separated in the angular direction
and have only local interactions.
We call this property an `asymptotic locality'.
This will play a crucial role in determining the scaling dimension of instanton as will be discussed later.

Now we can determine the scaling dimension of the instanton operator.
At a scale set by $\tau$ and $x$, 
fermion modes whose angular momenta are less than $n_{max} \sim k_F \mbox{min}( x, \tau ) = k_F x$ are twisted
(at sufficiently large distances, we always have $\tau > x$ because space has the smaller absolute scaling dimension).
In other words,
fermion fields that are twisted at scale $x$
have `wavelengths' larger than $\delta \theta \sim 1/n_{max} \sim ( k_F x)^{-1}$
in the space of $\theta$. 
Since $x$ and $\theta$ scale as $x = b^{2/3} x^{'}$ 
and $\theta = b^{-1/3} \theta^{'}$, we have
 $\delta \theta^{'} \sim b^{-1/3}  (k_F x^{'}) ^{-1}$.
Note that  $\delta \theta^{'} \rightarrow 0$ as $b \rightarrow \infty$ 
and fermion fields with arbitrarily small `wavelengths' are twisted in the low energy limit.
Since an instanton twists fermions of all angle, 
an instanton corresponds to an operator 
which creates a vortex with flux $\pi$ along the non-compact angle direction $\theta$.
The physical reason why an instanton which is localized in space and time becomes an extended object is as follows.
In the low energy limits, only small angle scatterings are important and 
fermions rarely change the directions of their motions.
They essentially move on 1+1D planes in space and time.
Since the distances from the instanton and the planes on which fermions move are fixed, 
at a sufficiently large distance scale, all the fermions acquire phase $\pi$ 
as they are transported around the instanton.
The 1+1D fermions are parametrized by the direction of their velocities, $\theta$
and an instanton becomes a vortex which is extended along the angular direction.

It is noted that even though the (Euclidean) Lorentz symmetry is broken by the non-relativistic fermions, 
the twist angle will be $\pi$ in the long distance limit
if there is the spatial rotational symmetry as discussed in Sec. III.
If there is no spatial rotational symmetry, the twist angle will depend on $\theta$.
But an argument similar to the one provided at the end of Sec. III can be made to extend the conclusion of the following discussion to more general cases.
In the following, we will focus on the rotationally symmetric case.

\begin{figure}
        \includegraphics[height=6cm,width=8cm]{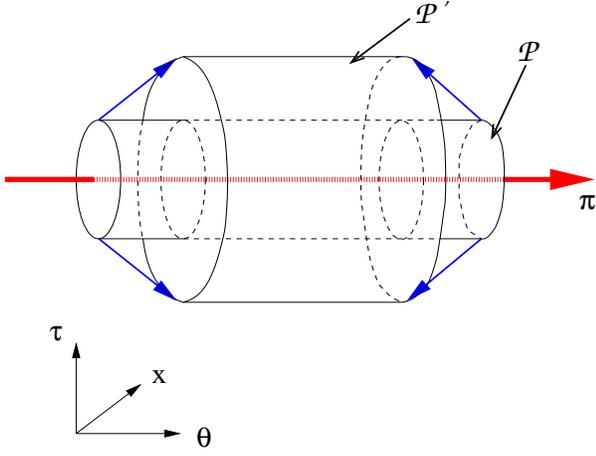} 
\caption{
`Time'-evolution of a quantum state defined on the surface of 
a pipe extended along the angular direction, where
a $\pi$-vortex is pierced through the pipe.
Under the time-evolution, a point on the surface ${\cal P}$ 
is mapped to a point on the surface ${\cal P^{'}}$. 
 }
\label{fig:scale}
\end{figure}

The scaling dimension of the extended twist operator can be obtained following 
the reasoning which is analogous to the state-operator correspondence in relativistic quantum field theories.
In relativistic cases, instanton corresponds to an operator defined at a point in space and time.
The operator defines a quantum state on the sphere enclosing the instanton operator.
The scaling dimension corresponds to the `energy' of the quantum state 
associated with the `time'-evolution in the radial direction.
In the present non-relativistic case, instanton corresponds to an extended vortex operator with flux $\pi$ 
because instanton twists boundary conditions of all fermions 
which are parametrized by the non-compact variable $\theta$.
The extended operator defines a quantum state on the surface of a pipe ${\cal P}$ 
which is extended in the $\theta$ direction in the space of $\tau$, $x$ and $\theta$ 
as is shown in Fig. \ref{fig:scale}.
In the functional Schrodinger picture, the vortex operator defines a quantum state as
\bqa
&& \left. \Psi[ \psi_{js}^{'}(\tau,x,\theta), \psi_{js}^{*'}(\tau,x,\theta), a^{'}_q(\tau,x, \theta) ] \right|_{ (\tau,x,\theta) \in {\cal P}} 
\nn && 
= 
\int D \psi_{js} D \psi_{js}^* D a_q e^{-S[\psi,\psi^*,a]}.
\eqa
On the r.h.s. of the above equation, the fermion fields have the anti-periodic boundary condition 
and all the fields inside the pipe are integrated out with the condition that the fields on the surface of the pipe
coincides with the fields $\psi_{js}^{'}(\tau,x,\theta), \psi_{js}^{*'}(\tau,x,\theta), a_q^{'}(\tau,x, \theta)$.
The scaling dimension corresponds to the `energy' of this quantum state associated with the time evolution 
given by 
\bqa
x & \rightarrow & b^{2/3} x, \nn
\tau & \rightarrow & b^{1} \tau, \nn
\theta & \rightarrow & b^{-1/3} \theta. 
\label{scale4}
\eqa
We can define a `Hamiltonian' for the `time' evolution 
because the action is local in $\tau$ and $x$.
Namely, a quantum state on a surface  
with larger $|\tau^{'}|$ and $|x^{'}|$ are uniquely determined from 
the state on a surface with smaller $|\tau|$ and $|x|$.

Since $\tau$ and $x$ scale differently, 
the surface of the pipe should `expand' in different rates 
depending on its normal vector.
Note that the `time' evolution also involves the transformation in $\theta$  
because of the anomalous dimension of the angular variable.
In determining the scaling dimension of instanton, 
the key is the locality of the action in the angular variable $\theta$.
The action Eq. (\ref{56}) is asymptotically local in the space of $\theta$ in the following senses.
First, angles of the fermions can change at most 
by $\tilde \Lambda / k_F$ through an interaction with the gauge field.
Second, the fermion field at angle $\theta$ is coupled only with the gauge field near angle $\theta$.
Third, two gauge fields which have different angles $\theta_1$ and $\theta_2$ are independent for a sufficiently large $|\theta_2 - \theta_1|$ which is, yet, much smaller than the range of the $\theta$.
One may worry about a possible breakdown of the locality 
in the angular direction in the presence of 
short range four fermion interactions.
Indeed, the four fermion interaction 
\bqa
&& V \int \Pi_{i=1}^4 d \omega_i dk_i d\theta_i ~~
\psi^*(\omega_1,k_1,\theta_1) 
\psi^*(\omega_3,k_3,\theta_3)  \nn
&& \psi(\omega_2,k_2,\theta_2) 
\psi(\omega_4,k_4,\theta_4) 
\delta( \sum_i (-1)^i  \omega_i ) \nn
&&
\delta(  \sum_i (-1)^i (k_F + k_i ) \cos \theta_i )
\delta(  \sum_i (-1)^i (k_F + k_i ) \sin \theta_i ) \nn
\eqa
is non-local in the angular direction.
However, the interaction strength scales as 
$V \rightarrow b^{-2/3} V$ under the transformations
in Eqs. (\ref{scale2}) and (\ref{scale3})
and the four fermion interaction is irrelevant at low energies.

Because of this locality of the action, 
any extended object should have an `energy' which is either zero or infinite with respect to the vacuum.
This is a very general statement for a local theory.
For example, if a local theory is defined in a range $0 < \theta < 2L$ 
where $L$ is much larger than any length scale of the local coupling in the theory,
the energy of the system is roughly the twice of the energy of the system defined in the range $0< \theta < L$.
This implies that only $0$ or $\pm \infty$ are possible for the energy of the vortex operator 
with respect to the vacuum energy.
The vortex is a non-trivial object which necessarily `excites' the fermionic state by 
twisting the boundary condition and it should have an infinite `energy'.
Note that $-\infty$ ($0$) are excluded because a negative (zero) energy would imply
that instanton becomes more relevant (equally relevant) with increasing number of spinon flavors
or with increasing length of Fermi surface in the momentum space.
This is unlikely because the fermions always screen the gauge field and 
the scaling dimension of instanton should increase as the number of fermion modes increases.
Although the scaling dimension defined in the thermodynamic limit is infinite, 
the scaling dimension of the regularized instanton operator 
is well defined and systematically increases as the number of available fermion modes increases.
The number of fermion modes in a finite system is
proportional to the number of spinon flavors and the length of the Fermi surface 
due to a finite mesh in the momentum space. 
This implies that the scaling dimension of the instanton operator diverges more rapidly
as the long distance limit is taken when there are more flavors or longer Fermi surface. 
An explicit example of this is provided in Eqs. (\ref{D}) and (\ref{D2}) for the non-interacting system,
where the scaling dimension of the regularized instanton operator increases as $N$ or $k_F$ increases.
Therefore we conclude that instanton has to have a positive infinite scaling dimension in general.

\section{Strongly interacting fermions with a few flavors}
Now we move on to the physical case with a small but nonzero $N$, e.g., $N=2$.
In this case, even if one can somehow ignore instantons, 
the fermions are already strongly coupled with the non-compact gauge field.
Therefore, one can not exclude the possibility that the strongly interacting fixed point 
becomes unstable against a particle-hole or particle-particle condensation\cite{ALTSHULER,Lee_amp,Galitski} 
due to the strongly fluctuating non-compact gauge field.
Here we set aside those possibilities caused by the non-compact gauge field 
and focus on the question whether the fixed point is stable against 
proliferation of instantons or not.
Because we can not ignore vertex corrections any more, 
we do not know what the precise form of the scale transformation is for the strongly interacting fixed point.
However, we can still argue that the scaling dimension of an instanton is infinite based on 
the locality of the low energy theory in the angular direction.
The emergence of the locality in the angular direction is independent of a particular form of scale transformation.
Because of the Fermi surface geometry, 
momentum of the gauge field should be scaled down more slowly
compared to radial momentum of the fermions. 
If a momentum of fermion scales as $k = b^{-a} k^{'}$ then
a momentum of the gauge field should scale as $q = b^{-a/2} q^{'}$,
which is the consequence of the fact that Fermi surface is locally parabolic unless
there is a nesting which we do not consider here.
Therefore, we have $q/k >>1$ at low energies 
and this forces fermions at a certain angle are coupled only with 
the gauge field whose momentum is tangent to the Fermi surface.
This guarantees that the low energy effective action should be local in the angular direction.
Moreover, the mixing between $k$ and $q$ gives rise to the anomalous scaling dimension
for the angular variable, $a/2 > 0$ and the angular variable becomes decompactified in the low energy limit.
Again, this implies that the instanton operator is an extended vortex with flux $\pi$ 
along the extended angular direction.
Because of the locality, the extended vortex should have an infinite scaling dimension as we have discussed previously.

\section{Conclusion}
In conclusion, we argue that the U(1) spin liquid state is stable against proliferation of instantons for any nonzero number of spinon flavors if there is a spinon Fermi surface.
We formulated the low energy modes near the Fermi surface in terms of an infinite set of chiral fermions
and made an observation that the angular variable that parametrize the Fermi surface acquires 
a positive scaling dimension due to quantum effects and becomes a non-compact variable in the low energy limit.
Because the low energy effective theory is local in the non-compact angular direction, an instanton, 
which twists boundary conditions of all chiral fermions, should have an infinite scaling dimension.
Therefore, instantons are strongly irrelevant and the non-compact U(1) gauge theory is a good low energy description.

A few comments are in order.
First, since the scaling dimension of instanton is infinite in the infrared limit, 
a small fugacity of instantons will rapidly flow to zero at low energies.
A finite fugacity at an intermediate length scale implies that there is a small but finite density of instantons at the length scale.
However, the renormalization group flow implies that the density at a larger distance scale 
will become smaller and instantons are not important in the long distance limit.
If the fugacity $y$ is tuned to a sufficiently large value via tuning some microscopic parameters, 
then the nonlinear terms in the flow equation, Eq. (\ref{flow}) becomes important and the sum of them may not converge.
This signifies a breakdown of the perturbative expansion which is valid within a finite domain near the $y=0$ point.
If this happens, the fugacity of the  instanton operator can flow toward a large value leading to a confinement,
despite the fact that the linear term has a large negative coefficient.
It is noted that the infinite scaling dimension of instanton at the deconfined fixed point $y=0$ 
does not necessarily imply that the confinement phase is always unstable.
Both the deconfinement phase and the confinement phase may have finite regions of stability in the parameter space, separated by a phase transition.
To understand the nature of the confinement phase is an important open problem\cite{SUBIR}.
Second, we expect that the present argument can be  generalized 
to the cases where there is no rotational symmetry or there are only  
segments of Fermi surface.
This has been already demonstrated for the free fixed point at the end of Sec. III.
We expect that the similar argument will hold true at the interacting fixed point too.
This is because any finite segment of Fermi surface contains
an infinite number of modes which contribute to the scaling dimension
of instanton.

\section{Acknowledgment}
This work has been supported by NSERC.
The author thanks Matthew Fisher, Yong Baek Kim, Subir Sachdev and Xiao-Gang Wen 
for helpful discussions and, 
particularly, Patrick Lee for helpful comments and suggestions to improve the paper.

\begin{figure}[t!]
        \includegraphics[height=7cm,width=7cm]{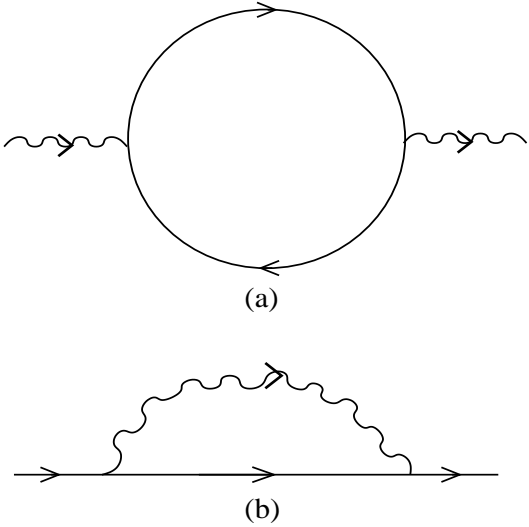} 
\caption{ 
Feynman diagrams for the one-loop self energies of (a) the gauge field and (b) the fermions. 
The solid (wiggled) line represents the propagator of the fermions (gauge field).
}
\label{fig:self}
\end{figure}

\renewcommand{\theequation}{A\arabic{equation}}
\setcounter{equation}{0}

\section*{APPENDIX A: Scaling dimension of a general twist operator}

Here we calculate the scaling dimension of a generalized twist operator.
Suppose the boundary condition of a fermion is twisted by an arbitrary angle around the twist operator
which is inserted at the origin.
The fermion satisfies the twisted boundary condition
\bqa
\psi(z e^{2\pi i} ) = e^{ i \Phi } \psi(z)
\eqa
in the complex plane with $z = \tau - i x$, where
$\Phi = 2\pi (1/2 - \eta )$ with $0 \leq \eta < 1$.
$\eta = 1/2$ corresponds to the untwisted periodic boundary condition
and $\eta  = 0$, the antiperiodic boundary condition.
A general value of $\eta$ between $0$ and $1$ corresponds to a fermion twisted
by a phase angle between $-\pi$ and $\pi$.
The twisted boundary condition is implemented by the mode expansion,
\bqa
\psi(z)  & =& \sum_{n} \frac{\psi^{n+\eta}}{z^{n+\eta+1/2}}, \nn
\psi^*(z)& =& \sum_{n} \frac{\psi^{*n-\eta}}{z^{n-\eta+1/2}},
\eqa
where $\psi^{n+\eta}$ and $\psi^{* n-\eta}$ with integer $n$ are the $n$-th normal modes. 
Note that $\psi$ and $\psi^*$ have the different mode expansions because
the gauge invariant operator $\psi^* \psi$ always have to satisfy the untwisted 
periodic boundary condition.
The scaling dimension of the twist operator is given by
\bqa
d & = & \left< \oint \frac{dz}{2\pi i}~~ z T(z) \right>_{twisted ~b.c.},
\label{a3}
\eqa
where the contour of the integration encloses the origin and
\bqa
T(z) = \frac{1}{2} : \left[ (\partial \psi^*) \psi - \psi^* \partial \psi \right] :
\label{a4}
\eqa
is the regularized holomorphic energy-momentum tensor 
with 
\bq
: \psi^*(z) \partial \psi(z): \equiv \lim_{w \rightarrow z} \left[ \psi^*(w) \partial \psi(z) + \frac{1}{(w-z)^2} \right].
\label{a5}
\eq
To calculate $ < T(z) >_{twisted~ b.c.}$, we first consider the expectation value
of a fermion bilinear with the twisted boundary condition,
\bqa
 && \left< \psi^*(w) \psi(z) \right>_{twisted~ b.c.} \nn
&& = \sum_{m,n} \left< \frac{ \psi^{m-\eta *} }{w^{m+1/2-\eta}}  \frac{ \psi^{n+\eta} }{z^{n+1/2+\eta}} \right> \nn
&& = \frac{ w^{\eta-1/2} z^{-\eta+1/2}}{w-z}, 
\label{a6}
\eqa
where we have used the commutator $\{ \psi^{m-\eta \dagger}, \psi^{n+\eta} \} = \delta_{m,-n}$
and the property of the vacuum, $  \psi^{n+\eta} |0> = \psi^{n-\eta \dagger} |0> = 0$ for $n>0$\cite{POLCHINSKI_BOOK}.
From Eqs. (\ref{a4})-(\ref{a6}), we obtain
\bqa
< T(z) >_{twisted ~ b.c.} = \frac{ ( 2 \eta - 1)^2}{8 z^2}
\eqa
and from Eq. (\ref{a3}) the scaling dimension is obtained to be $\frac{( 2\eta - 1)^2}{8}$.
In the presence of multiple fermions,
the total energy momentum tensor is 
the sum of  individual energy momentum tensors and
the scaling dimension is the sum of individual scaling dimensions.
If the $i$-th fermion is twisted by an angle $\Phi_i$,
the scaling dimension of the twist operator becomes
\bqa
d = \sum_{i} \frac{ \Phi_i^2 }{8 \pi^2}.
\eqa

\renewcommand{\theequation}{B\arabic{equation}}
\setcounter{equation}{0}

\section*{APPENDIX B: Calculation of the self energies}

In the large $N$ limit, the vertex corrections are negligible and the one-loop corrections are dominant\cite{POLCHINSKY}.
Although the one-loop self energies of the fermions and the gauge field are well known\cite{POLCHINSKY,KIM94,ALTSHULER},
it is instructive to calculate them from the effective action Eqs. (\ref{53})-(\ref{54})
to make sure that the theory contains the essential low energy physics.

From the quadratic action in Eq. (\ref{53}), the propagator of the fermion is given by
\bqa
&&\left< \psi_{js}(\omega,k,\theta) \psi_{j^{'}s^{'}}^*(\omega^{'},k^{'},\theta^{'} ) \right> 
 =  \delta_{j,j^{'}} \delta_{s,s^{'}}  \nn
&& \delta( \omega-\omega^{'}) \delta(k-k^{'}) \delta( \theta - \theta^{'} ) g_s( \omega , k, \theta ),
\eqa
where $g_s( \omega , k, \theta ) = \frac{1}{ i \omega + s k}$,
and the propagator of the gauge field is given by
\bqa
\left< a(\nu,q,\alpha) a(\nu^{'},q^{'},\alpha^{'} ) \right> 
& = & \frac{1}{|q|} \delta( \nu+\nu^{'}) \delta(q+q^{'}) \delta(\alpha - \alpha^{'}) \nn
&& {\cal D}_0( \omega , q, \alpha ),
\eqa
where ${\cal D}_0( \nu , q, \alpha ) = \frac{1}{ \frac{1}{g^2}(\nu^2 + q^2) + 2K }$.
The unconventional factor, $1/|q|$ in the gauge propagator is due to the angular representation.

Applying the standard Feynman rule to the action in Eq. (\ref{54}),
we can calculate the  self energy of the gauge field (Fig. \ref{fig:self} (a)) as
\bqa
&& \Pi(\nu,q,\alpha)  =  \frac{N k_F}{(2\pi)^3} \sum_s \int d\omega dk d\phi   \nn
&& g_s( \omega + \nu/2, k+q\phi/2, \phi + \alpha + sq/2k_F ) \nn
&& g_s( \omega - \nu/2, k-q\phi/2, \phi + \alpha - sq/2k_F ) \nn
&& = -\frac{N k_F}{2 \pi} + \frac{N k_F}{4\pi} \frac{|\nu|}{|q|}
\eqa
for $|q| >> |\nu|$.
The dressed gauge propagator becomes 
\bqa
{\cal D}^{-1}(\nu, q, \theta-\phi) = \left( 2 K - \frac{N k_F}{2 \pi}  \right) + \gamma \frac{|\nu|}{|q|} + \chi q^2,
\eqa
where $\gamma$ and $\chi$ are constants.
The first term should be canceled due to the gauge invariance.
The diamagnetic term $K$ depends on the details of the high energy cut-off (regularization) scheme
and we fix its value by requiring the gauge invariance.
From Fig. \ref{fig:self} (b), we obtain the self energy of the fermions as
\bqa
&& \Sigma_s(\omega,k,\theta) = -\frac{1}{(2 \pi)^3} \int d\nu dq |q| d\phi \nn
&& {\cal D}(\nu, q, \theta-\phi) g_s( \omega-\nu, k-q \phi, \theta - sq/k_F), 
\eqa
where it is essential to use the dressed gauge propagator,
\bqa
{\cal D}(\nu, q, \alpha) = \frac{1}{ \gamma |\nu|/|q| + \chi q^2 }.
\eqa
The integration can be done straightforwardly and we obtain,
\bq
\Sigma_s(\omega,k,\theta) = i c~ \mbox{sgn}(\omega) |\omega|^{2/3},
\eq
where $c \sim \gamma^{-1/3} \chi^{-2/3}$ is a constant.
Even though we use the dressed fermion propagator, 
the leading behaviors of the self energies are not modified\cite{POLCHINSKY}.


\begin{thebibliography}{99}
\bibitem{NONLOCAL} The fact that the twist angles saturate to $\pi$ only in the large $L$ limit is due to the non-local nature of the instanton operator.
\bibitem{GAUGE} It is convenient to choose the temporal gauge for the singular part and the Coulomb gauge for the non-singular part. Note that it is possible to choose different gauges for singular and non-sinular parts.
\end{thebibliography}
\end{document}